\documentclass[aps,prd,twocolumn,superscriptaddress,nofootinbib,floatfix]{revtex4-1}
\usepackage{graphicx,bm,dcolumn,color,epsfig,amsmath,amssymb,rotating,braket,float,comment}
\newcommand{\non}{\nonumber}
\newcommand{\ba}{\begin{eqnarray}}
\newcommand{\ea}{\end{eqnarray}}
\newcommand{\D}{\nabla}

\begin{document}

\title{Gravitational radiation in Infinite Derivative Gravity \\
and connections to Effective Quantum Gravity}
\author{James Edholm}
\affiliation{Physics Department, Lancaster University, Lancaster, LA1 4YW}

\begin{abstract}
The Hulse-Taylor binary provides possibly the best test of GR to date. We  find the modified quadrupole formula for Infinite Derivative Gravity (IDG). 
We investigate the backreaction formula for propagation of gravitational waves, found previously for Effective Quantum Gravity (EQG) for a flat background and extend this calculation to a de Sitter background for both EQG and IDG. We put tighter constraints
on EQG using new LIGO data.
We also find the power emitted by a binary system 
within the IDG framework for both circular and elliptical orbits 
and use the example of the Hulse-Taylor binary to show that IDG is consistent with GR. 
\end{abstract}
\maketitle
%https://arxiv.org/pdf/1104.2200.pdf
%Gravitational radiation in f(R) gravity

General Relativity (GR) has been spectacularly successful in experimental tests,
notably in the recent detection of gravitational waves \cite{Abbott:2016blz}. 
One of the most renowned tests is the Hulse-Taylor binary. 
The way the orbital period of these two stars 
changes over time depends on the gravitational radiation emitted. 
This matches the GR prediction to within 0.2\%~\cite{Weisberg:2016jye}. 
 
However, GR breaks down at short distances where it produces 
singularities. The first attempts to modify gravity by altering the action
failed because they generated ghosts, which are excitations
with negative kinetic energy~\cite{stelle:1977}. Infinite Derivative Gravity (IDG)
\cite{Tseytlin:1995uq,Biswas:2005qr,Biswas:2011ar,Biswas:2016etb,Siegel:2003vt,Biswas:2005qr,
Biswas:2011ar,Biswas:2016etb,Buoninfante:2016iuf,Talaganis:2014ida,Modesto:2011kw,Modesto:2012ys,
Biswas:2005qr,Biswas:2011ar,
Edholm:2016hbt,Conroy:2017nkc,Edholm:2018dsf,Conroy:2014eja,Cornell:2017irh,Biswas:2013cha,
Calcagni:2013vra,Biswas:2010zk,Biswas:2012bp,Koshelev:2012qn,Koshelev:2013lfm,Biswas:2016etb,Biswas:2016egy,
Conroy:2015wfa,
Edholm:2016seu,Briscese:2012ys,Teimouri:2016ulk,Koshelev:2016xqb,Craps:2014wga,Talaganis:2014ida,
Talaganis:2017tnr,Conroy:2014dja} 
 avoids this fate 
while also allowing us the possibility to not produce singularities.

IDG has the action \cite{Biswas:2011ar}
\ba \label{IDGactionweyl}
        \mathcal{L}= \frac{\sqrt{-g}}{2}\bigg[&& M^2_P R\ + R F_1 (\Box) R + R_{\mu\nu} F_2(\Box) R^{\mu\nu}
       \non\\
       &&+ C_{\mu\nu\rho\lambda} F_3(\Box) C^{\mu\nu\rho\lambda}\bigg],
\ea
where $M_P$ is the Planck mass, $R$ is the Ricci scalar, $R_{\mu\nu}$ is the Ricci tensor and 
$C_{\mu\nu\rho\lambda}$ is the Weyl tensor. Each $F_i(\Box)$ is an infinite series of the d'Alembertian operator $\Box=g^{\mu\nu}\nabla_\mu \nabla_\nu$
i.e. $F_i(\Box)=\sum_{n=0}^\infty f_{i_n} \Box^n/M^{2n}$, where the $f_{i_n}$s are dimensionless coefficients
and $M$ is the mass scale of the theory, which dictates the length scales below which the additional terms 
come into play. 

The propagator $\Pi_{\text{IDG}}$ around a flat background in terms of the spin projection operators is modified as follows \cite{Biswas:2011ar}
\ba
        \Pi_{\text{IDG}} = \frac{P^2}{a(k^2)}+\frac{P^0_s}{a(k^2)-3c(k^2)}\underset{a=c}{=}
        \frac{\Pi_{\text{GR}}}{a(k^2)}
\ea
where $a$ and $c$ (given in \eqref{eq:defnacfforflat}) are combinations of the $F_i(\Box)$s from \eqref{IDGactionweyl}. In the second equality we have taken the simplest choice $a(k^2)=c(k^2)$, giving a clear path back to GR in the limit $a(k^2)\to 1$. 

The simplest way to show that there are no ghosts is to show that there are no poles in the propagator, which means there can be no zeroes in
$a(k^2)$. Any function with no zeroes can be written in the form of the 
exponential of an entire function, so we choose 
$a(k^2)=c(k^2)=\exp\left[\gamma(k^2/M^2)\right]$, where $\gamma$ is an entire function.   

Any entire function can be written as a polynomial 
$\gamma(k^2)=c_0+c_1k^2+c_2k^4+\cdots$, so a priori 
we have an infinite number of coefficients to choose. 
However, it was shown that only the first few orders will appreciably affect the predictions of the theory, as terms higher than order $\sim 10$ can be described
by a rectangle function with a single unknown parameter \cite{Edholm:2018wjh}.

The quadrupole formula tells us the perturbation 
to a flat metric caused by a source with quadrupole moment $I_{ij}$.
Here we use the equations of motion to find the modified quadrupole formula for IDG.

\section{Modified quadrupole formula}
The IDG equations of motion for a perturbation $h_{\mu\nu}$ around a flat background $\eta_{\mu\nu}$ 
are given by \cite{Biswas:2011ar}
\begin{widetext}
\ba
        -\kappa T_{\mu\nu} = \frac{1}{2} \bigg[a(\Box)\left(\Box h_{\mu\nu} 
        -\partial_\sigma
        \left(\partial_\mu h^\sigma_\nu + \partial_\nu h^\sigma_\mu \right)\right)
        + c(\Box) \left(\partial_\mu \partial_\nu h + \eta_{\mu\nu} \partial_\sigma \partial_\tau 
        h^{\sigma\tau}-\eta_{\mu\nu} \Box h\right)+ f(\Box) \partial_\mu \partial_\nu 
        \partial_\sigma \partial_\tau 
        h^{\sigma\tau}\bigg],
\ea        
\end{widetext}
where $\kappa=M^{-2}_P$ and 
\ba \label{eq:defnacfforflat}
        a(\Box) &=& 1 + M^{-2}_P \left(F_2(\Box) 
        + 2  F_3(\Box)\right) \Box, \non\\
        c(\Box) &=& 1 - M^{-2}_P\left(4 F_1(\Box) -  F_2(\Box)  
        + \frac{2}{3}F_3(\Box)\right)\Box,\non\\
        f(\Box) &=& M^{-2}_P \left(4F_1(\Box) + 2F_2(\Box) +\frac{4}{3} F_3(\Box)\right),
\ea        
and it should be noted that as $a(\Box)=c(\Box)$, then $f(\Box)\Box=a(\Box)-c(\Box)=0$. If we take the de Donder gauge $\partial_\mu h^{\mu\nu} = \frac{1}{2} \partial^\nu h$ and assume
$a(\Box)=c(\Box)$, then 
\ba \label{eq:linaeriseomsinddgauge}
        -2\kappa T_{\mu\nu} 
        = a(\Box)\Box \bar{h}_{\mu\nu}, 
\ea
where we have defined $\bar{h}_{\mu\nu} \equiv h_{\mu\nu}   - \frac{1}{2}g_{\mu\nu} h$ 
\footnote{Alternatively, we can follow the method of \cite{Naf:2011za} and define the
gauge $\partial^\mu \gamma_{\mu\nu}=0$,where \\$\gamma_{\mu\nu} = a(\Box) h_{\mu\nu} - \frac{1}{2}\eta_{\mu\nu}  
 c(\Box) h -\frac{1}{2}\eta_{\mu\nu} f(\Box) \partial_\alpha \partial_\beta h^{\alpha\beta}$. 
 This produces the result $-2\kappa T_{\mu\nu}=\Box \gamma_{\mu\nu}$.}. 
Note that in the limit $a\to 1$, we return to the GR result. 
We invert $a(\Box)$ and follow the usual GR method \cite{Carroll:2004st} where we assume the source 
is far away, composed of non-relativistic matter and isolated. In this approximation, the Fourier transform of $h_{\mu\nu}$ with 
respect to time is
\ba
        \tilde{\bar{h}}_{\mu\nu} = 4 G \frac{e^{i k r}}{r} 
        \int d^3 y \frac{\tilde{T}_{\mu\nu}(k,y)}{a(k^2)}.
\ea
 When we insert the definition of the quadrupole moment, 
 $I_{ij}=\int d^3 y~ T^{00}(y) y^i y^j$, 
write out the full expression for $\tilde{I}_{ij}$ 
and define the retarded time $t_r=t-r$, we obtain
\ba \label{eq:finaleqnwithgenerala}
        \bar{h}_{ij} &=& \frac{ - G }{\pi} \frac{1}{r}\frac{d^2}{dt^2}\int dk dt'_r  
        \frac{e^{ik(t_r-t'_r)}}{a(k^2)} I_{ij} (t'_r).
\ea

\section{Simplest choice of $a(\Box)$}
We choose $a(k^2)$ to avoid ghosts, by ensuring there are no poles in the propagator.
If we choose $a(k^2)=e^{k^2/M^2}$ and use the formula for the inverse Fourier transform of a Gaussian,
we find
\ba
        \bar{h}_{ij} &=&    \frac{- G}{r}\frac{M}{\sqrt{\pi}}\frac{d^2}{dt^2}\int  dt'_r  
         e^{-M^2(t_r-t'_r)^2/4} I_{ij} (t'_r).
\ea
This is the modified quadrupole formula for the simplest case of IDG.
We now need to specify $I_{ij}$. For example, 
when we look at the radiation emitted by a binary system of stars of mass 
$M_s$ in a circular orbit,
the 11 component of $I_{ij}$ is $I_{11}(t)=M_s R^2 \left(1+\cos(2\omega t)\right)$,
where $R$\ is the distance between the stars and $\omega$ is their angular velocity. 
Therefore         
\ba
        \bar{h}_{11} 
         =   \frac{4GM_s^2R^2}{r}\left(1+e^{-\frac{4\omega^2}{M^2}} 
         \cos (2\omega t_r)\right),
\ea
Comparing to the GR case, we see that this matches the GR prediction at large $M$, but at small 
$M$ there is a reduction in the magnitude of the oscillating term compared to GR.

\section{Backreaction equation}
There is a second order effect where gravity couples to itself 
and produces a backreaction.
In \cite{Kuntz:2017pjd}, the backreaction was found for Effective Quantum Gravity (EQG). 
EQG has a similar action to IDG (the $F_i(\Box)$
in \eqref{IDGactionweyl} are replaced by $a_i + b_i \log(\Box/\mu^2)$ where $\mu$ is a 
mass scale \cite{Donoghue:2012zc,Calmet:2017qqa,Calmet:2018qwg}.

In this section we generalise the result of \cite{Kuntz:2017pjd} 
(see also \cite{Stein:2010pn,Preston:2016sip,Saito:2012xa}) and 
also extend it to a de Sitter background.
Using the Gauss-Bonnet identity and a similar expression for the 
higher-order terms \cite{Li:2015bqa} we can focus on
\eqref{IDGactionweyl} without the Weyl term.

Far away from the source, we use the gauge $\D^\mu h^\nu_\mu= 0$ and $h=0$,
to simplify the linearised and quadratic (in $h_{\mu\nu}$) curvatures around a de Sitter background, 
given in 
\eqref{eq:desitterperturbedcurvatures} and \eqref{eq:desitterquadperturbedcurvatures}.

The linear vacuum equations of motion around a dS background in this gauge \cite{Conroy:2017uds,Edholm:2017fmw} are 
\begin{widetext}
\ba \label{eq:linearvacuumdseqnsbackground}
        (\Box-2H^2)^2F_2(\Box)h^\mu_\nu = -\left(1+24M^{-2}_P H^2 f_{1_0}\right)\left(\Box-2H^2
\right)h^\mu_\nu,
\ea
\end{widetext}
where $f_{1_0}$ is the zeroeth order coefficient of $F_1(\Box)$ and
the background Ricci curvature scalar is $\bar{R}=12H^2$, where $H$ is the Hubble constant.
Upon inserting \eqref{eq:linearvacuumdseqnsbackground}
into the averaged second order equations of motion for the 
non-GR terms \eqref{eq:averagedquadraticeoms},
\ba \label{eq:fullbackreactioneqn}
        \kappa t^\mu_\nu{}^{\text{IDG}} 
        &=&\left(1+24M^{-2}_P H^2 f_{1_0}\right)\bigg[
        -\frac{1}{2} \braket{ h^\mu_\sigma\left(\Box-2H^2
        \right)h^\sigma_\nu}\non\\
        &&+\frac{1}{8}\delta^\mu_\nu \braket{ h^\tau_\sigma \left(\Box-2H^2
\right)h^\sigma_\tau}\bigg], 
\ea         
%(note that the contraction of this vanishes). 
where $f_{1_0}$ corresponds to $b_1$ in the EQG formalism 
and $\left<X\right>$ represents the spacetime average of $X$ using the 
same definition as \cite{Kuntz:2017pjd}. 

\eqref{eq:fullbackreactioneqn} is the full backreaction equation for any action with higher derivative terms which is quadratic in the curvature; we have not used the fact that IDG contains an infinite series of the d'Alembertian
and so this method can be applied to finite higher derivative actions, for example \cite{Giacchini:2018gxp,Boos:2018bhd}.

So the energy density $\rho=t_{00}$ is given by 
\ba
        \rho^{\text{IDG}}_{dS} &=&  \left(M^2_P+24 H^2 f_{1_0}\right)\bigg[ 
        \frac{1}{2} \braket{ h_{0\sigma}\left(\Box-2H^2
        \right)h^\sigma_0}\non\\
        &&+\frac{1}{8} \braket{ h^\tau_\sigma \left(\Box-2H^2
\right)h^\sigma_\tau}\bigg].
\ea

For a plane wave\footnote{There are extra terms due to the 
de Sitter background  
$H^2$ \cite{Nowakowski:2008de,Arraut:2012xr}, but to linear order in, these produce
only terms which are linear is $\cos$ or $\sin$. The spacetime 
average therefore vanishes and there is no extra contribution to  
\eqref{eq:dsdensity} from these terms.} solution
 $h_{\mu\nu} = \epsilon_{\mu\nu} \cos(\omega t-kz)$,
we find (including the GR term)
\ba \label{eq:dsdensity}
        \rho_{dS} &=&\frac{1}{4}M^2_P\left(1 +24M^{-2}_PH^2 f_{1_0}\right)\bigg\{\omega^2\epsilon^2\non\\
        &&+2\left( 4 \epsilon^\sigma_0
        \epsilon_{0\sigma}+\epsilon^2  \right)\left(8H^2+ \omega^2-k^2\right)\bigg\},
\ea
where $\epsilon^2 = \epsilon_\mu \epsilon^\mu$. Given the current value of the Hubble constant $H_0=$, $H_0^2 M_P^{-2}\approx 10^{-119}$. Therefore $f_{1_0}$ would have to be of the order of
$10^{115}$ for the de Sitter background in the present day to have a noticeable impact. Thus we can generally use the 
Minkowski background as a good approximation. In the EQG notation, $f_{1_0}$ is replaced by $b_1$ which already has the constraint $b_1<10^{61}$ so we can ignore this extra term.

For a classical wave, $\omega^2=k^2$ so the term on the second line of \eqref{eq:dsdensity}
disappears
for a Minkowski background.
This is the case for IDG  when we assume there are no extra poles in the propagator.
On the other hand, EQG does have poles, 
so for EQG or IDG with a single pole there can be damping 
\cite{Calmet:2016sba,Calmet:2014gya,Calmet:2017omb,Calmet:2016fsr} and therefore $\omega^2\neq k^2$.

Kuntz used LIGO constraints on the density parameter $\Omega_0$ 
as well as the constraint on the mass of the pole $m>5\times 10^{13}$ GeV 
to constrain $\epsilon$, the amplitude of the massive mode as $\epsilon<1.4 \times 10^{-33}$
\cite{Kuntz:2017pjd}.
Since then, LIGO has found more stringent constraints of $\Omega_0<5.58\times10^{-8}$
\cite{Abbott:2018utx}. Following the same method as \cite{Kuntz:2017pjd}, we divide by the 
critical density $\rho_c=\frac{3H^2_0}{8\pi G}$ to find
\ba
        \Omega_0 = \frac{1}{12} \left(\epsilon^\alpha_0 \epsilon_{0\alpha} +
        \epsilon^2\right) \frac{m^2}{H_0^2}<5.58\times10^{-8}
\ea
which we use to find a stronger
constraint of $\epsilon < 8.0 \times 10^{-34}$.
This cuts the allowed parameter space nearly in half and makes it less likely that the 
detector \cite{Baker:2009zzb} referred to in \cite{Kuntz:2017pjd} would be able to detect this mode.

\section{Power emitted}
We can use the backreaction equation 
to find the power radiated to infinity
by a system, which is given by \cite{Carroll:2004st}
\ba
        P=\int_{S_\infty^2} t_{0\mu} n^\mu r^2 d\Omega, 
\ea
where the integral is taken over a two-sphere at spatial infinity
$S_\infty^2$ and $n^\mu$ is the spacelike normal vector to the two-sphere.
In polar coordinates, $n^\mu=(0,1,0,0)$. We are therefore
 interested in the $t_{0r}$ component. 
 
In the limit $H \to 0$ and including the usual GR term,
\eqref{eq:fullbackreactioneqn} becomes
\ba \label{eq:backreactioneom}
        t_{\mu\nu}=\frac{1}{64\pi G}\bigg[&&2\braket{\partial_\mu h^{TT}_{\alpha\beta} \partial_\nu 
        h^{\alpha\beta}_{TT}}
        + 4 \braket{ h^{TT}_{\sigma(\mu} 
\Box_\eta h^{TT\sigma}_{\nu)}} \non\\
        &&-\eta_{\mu\nu}
\braket{h^{TT}_{\sigma\tau} \Box_\eta h_{TT}^{\tau\sigma}}\bigg].
\ea 

Note that $h^{TT}_{0\nu}=\eta_{0r}=0$, 
which means we can discard the second and third terms in the square bracket.
The relevant term for the power becomes 
\ba
        t_{0\mu}n^\mu = \frac{- GM^2}{32\pi^2 r^2}\left<\frac{d^3}{dt^3}\left(\hat{I}_{ij}(t_r)\right)
        \frac{d^3}{dt^3}\left(\hat{I}^{ij}(t_r)\right)\right>.
\ea        
Note that this is the same as the GR expression, but where we have defined 
$\hat{I}_{ij}=\int  dt'_r  
         e^{-M^2(t_r-t'_r)^2/4} I_{ij} (t'_r)$ instead of $I_{ij}$.
If we convert to the reduced quadrupole moment $\hat{J}_{ij}$, using $J_{ij}=I_{ij} - \delta_{ij} \delta^{kl} I_{kl}$ \cite{Carroll:2004st}, 
we can use the identities \eqref{eq:integralidentities} from \cite{Carroll:2004st}
to see that the power emitted by a system is
\ba \label{eq:generalradiationpowerformula}
        P = - \frac{G}{5} \left<\frac{d^3\hat{J}_{ij}}{dt^3}
        \frac{d^3\hat{J}^{ij}}{dt^3}\right>,
\ea
where $\hat{J}_{ij}= \int^\infty_{-\infty} dt'_r  
         e^{-M^2(t_r-t'_r)^2} J_{ij} (t'_r)$.
This result can then be applied to any system for which we know the reduced quadrupole
moment. We will now apply it to binary systems in both circular and elliptical orbits.      
\subsection{Circular orbits}
For a binary system of two stars in a circular orbit, the reduced quadrupole 
moment $J_{ij}$ in polar coordinates is given in \cite{Carroll:2004st} and depends on 
the mass of each of the stars $M_s$,  the distance between them $R$, and 
the angular velocity $\omega$.\footnote{The corrections to the orbital motion
due to the change in the Newtonian potential from IDG will be negligible as this
has already been constrained down to the micrometre scale, much shorter than the
distance between the stars.}
Using \eqref{eq:generalradiationpowerformula}, our power is  (again in the limit $r\to \infty$)      
and using $\left<\sin ^2(x)\right>\equiv\frac{1}{2}$, 
\ba
       P&=& - \frac{128}{5}  GR^2M_s^4 \omega^6 e^{-2 \omega ^2/M^2}. 
\ea  
This is the GR result with an extra factor of $e^{-2 \omega ^2/M^2}$ where $M$ 
is the IDG mass scale.
This gives a reduction in the amount of radiation emitted from a binary system of stars
in a circular orbit. Note that this factor tends to 1 in the GR
limit $M\to \infty$.

\subsection{Generalisation to elliptical orbits}
\begin{figure}[tbp]\hspace{-6mm}
\includegraphics[width=91mm]{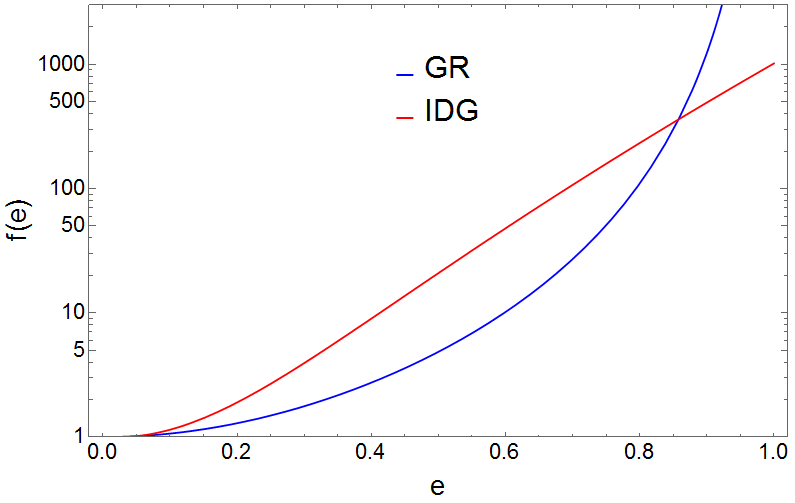}
\caption{The enhancement factor $f^{\text{IDG}}(e)$ given by \eqref{eq:fullidgenhancementfactor}
against the eccentricity $e$ as well as the enhancement factor for the GR term $f^{\text{GR}}(e)$, where the total
power is $P_{\text{GR}}^{\text{circ}} f^{\text{GR}}(e)+
        P_{\text{IDG}}^{\text{circ}} f^{\text{IDG}}(e)$. This factor describes
how the power emitted changes with respect to the eccentricity. The extra IDG term will show
up most strongly at around $e=0.7$, which coincidentally is close to the value for the Hulse-Taylor binary (0.617).}
\label{enhancementfactorplot}
\end{figure}

The power radiated by a binary system with a circular orbit is of limited 
applicability because in GR the power emitted is highly dependent 
on the eccentricity $e$ of the orbit~\cite{Peters:1963ux}, 
i.e. $P_{\text{GR}}=P_{\text{GR}}^{\text{circ}} f^{\text{GR}}(e)$.
where $ f^{\text{GR}}(e)$ is an \textit{enhancement factor} that reaches $10^3$ at $e=0.9$.
The circular orbit is therefore unlikely to be an accurate approximation.

For an elliptical orbit, the relevant components of the reduced quadrupole moment are \cite{Peters:1963ux}
\ba
        \hspace{-2mm}J_{xx}= \mu d^2 \left(\cos^2(\psi) - \frac{1}{3}\right),   \quad
        J_{yy}= \mu d^2 \left(\sin^2(\psi) - \frac{1}{3}\right),~~~~~~
\ea        
where $\mu$ is the reduced mass $m_1 m_2/(m_1 +m_2)$ and
the distance $d$ between the two bodies is given by        \\$d= \frac{a(1-e^2)}{1+e\cos(\psi)}$,
where $e$ is the eccentricity of the orbit and $a$ is the semimajor axis~\cite{Peters:1963ux}.
The change in angular position over time is       
\ba
        \dot{\psi}&=&\frac{\left[G(m_1+m_2)a(1-e^2)\right]^{1/2}}{d^2}.
\ea 
For the $xx$ component, we need to calculate
\vspace{4mm}
\ba \label{eq:xxcomponentfirstint}
        \hspace{-1mm}\hat{J}_{xx}=\mu a^2(1-e^2)^2 \int^\infty_{-\infty} dt'_r e^{-M^2 (t_r-t'_r)^2} 
        \frac{\cos^2(\psi( t'_r))-\frac{1}{3}}
        {\left(1+e\cos(\psi (t'_r))\right)^2}.~~~~~~
\ea

This is a very difficult integration to do. However, if we make the change of coordinates 
$z=M(t_r-t'_r)$, we can use a Taylor expansion in $\frac{1}{M}$ 
if it is small and the identities
\eqref{eq:evenoddidentities}
to see that we can write down \eqref{eq:expansionofhatJxx}, i.e. 
\ba \label{eq:totalpowerelliptical}
        P\approx P_{\text{GR}} + P_{\text{IDG}}=P_{\text{GR}}^{\text{circ}} f^{\text{GR}}(e)+
        P_{\text{IDG}}^{\text{circ}} f^{\text{IDG}}(e),~~~
\ea
where the IDG power for an elliptical orbit is the 
power for a circular orbit multiplied by an enhancement factor 
$f(e)$ which depends on the eccentricity.

\vspace{20mm}
We find that
\ba
        P_{\text{IDG}}=P_{\text{IDG}}^{\text{circ}} f^{\text{IDG}}(e)
        = \frac{256}{5}  \frac{\omega ^8}{M^2}GR^2M_s^4 f^{\text{IDG}}(e),~~~
\ea     
where $f^{\text{IDG}}(e)$ is a polynomial of 22nd order and so is given in the appendix. In the limit $M\to \infty$, $P_{\text{IDG}}\to 0$
and \eqref{eq:totalpowerelliptical} returns to $P_{\text{GR}}$.
$f^{\text{IDG}}(e)$ is plotted in Fig \ref{enhancementfactorplot} 
with a comparison to the enhancement factor for GR, $f^{\text{GR}}(e)$.

\interfootnotelinepenalty=10000
The  Hulse-Taylor binary has a period of 7.5 hours and 
ellipticity of 0.617.
The radiation emitted from the Hulse-Taylor binary is $0.998\pm 0.002$ 
of the GR prediction \cite{Weisberg:2016jye},
which leads to the constraint $M>6.9 \times 10^{-49} M_P= 1.0\times 10^{-21} \text{eV}$
on our mass scale $M$, which is much weaker than previous 
constraints. 

The previous lower bound \footnote{If we assume IDG is responsible for 
inflation we can obtain an even stronger lower bound of roughly $10^{14}$
GeV using Cosmic Microwave Background 
data \cite{Ade:2015xua,Edholm:2016seu,Koshelev:2016xqb}.}
is $\sim$0.01 eV from lab-based experiments \cite{Edholm:2016hbt}.
In order to produce a comparable constraint, we would need 
to study radiation produced from systems with orbital periods\footnote{The frequency
of the radiation produced
is twice the orbital frequency of the system \cite{Abbott:2016bqf}.} of
less than $10^{-4}$ seconds.
Not only do these systems have an orbital frequency much higher than
LIGO and LISA will be able to probe 
(15-150 Hz \cite{Abbott:2016bqf} and 
$10^{-4}$-$10^{-1}$\hspace{0.2mm}Hz \cite{Audley:2017drz} 
respectively), but they would also be out of the weak-field 
regime we used for our calculations.
Therefore
lab-based experiments and  CMB data are likely to provide the tightest constraints
in the near future. 
 
\section{Conclusion}
We found the modified quadrupole formula for IDG, which describes how 
the metric changes for a given stress-energy tensor. We generalised the backreaction formula 
already found for Effective Quantum Gravity (EQG) to a de Sitter background (for both EQG and IDG).
We used updated LIGO results to give a tighter constraint of 
 $\epsilon < 8.0 \times 10^{-34}$ on the 
amplitude of the massive mode in EQG.

Finally, we found the power emitted by a binary system, for 
both circular and elliptical orbits and investigated the example of the Hulse-Taylor binary. 
We showed that IDG is consistent with the GR predictions.

\iffalse
LIGO is most sensitive to systems with orbital
frequency of around 15-150 Hz \cite{Abbott:2016bqf} , 
and LISA will probe even lower frequencies ($10^{-4}$-$10^{-1}$\hspace{0.2mm}Hz) 
\cite{Audley:2017drz}. 
\fi

\vspace{9mm}
\section{Acknowledgements}
We would like to thank David\ Burton, Iber\^e Kuntz and Sonali Mohapatra
for their help in preparing this paper. 

JE is funded by the Lancaster University Faculty of Science and Technology.
\newpage
\appendix
\section{Linearised and quadratic curvatures}
The linearised Ricci curvatures around a de Sitter
background are \cite{Conroy:2017uds}
\ba \label{eq:desitterperturbedcurvatures}
        r^\mu_\nu &=& H^2 h^\mu_\nu
        - \frac{1}{2} \Box h^\mu_\nu, \quad \quad r= 0.~~~~~~
\ea
The curvatures to quadratic order are 
\ba \label{eq:desitterquadperturbedcurvatures}
        r^{(2)}_{\mu\nu}&=&\frac{1}{4}\left( h^{\alpha\beta} \nabla_\mu\nabla_\nu h_{\alpha\beta}
        -2h_{\alpha(\nu} (\Box-4H^2)h^\alpha_{\mu)}\right)  \non\\
        r^{(2)}&=& -\frac{1}{4}h_{\mu\nu} \left(\Box-8H^2\right)h^{\mu\nu}.
\ea    
The averaged second order equations of motion are 
\ba \label{eq:averagedquadraticeoms}
        \kappa t^\mu_\nu{}^{\text{IDG}} &=& \frac{1}{2} \braket{h^\mu_\sigma F_2(\Box)\left(\Box-2H^2
\right)^2h^\sigma_\nu} \non\\
        &&-\frac{1}{8}\delta^\mu_\nu
\braket{h^\tau_\sigma F_2(\Box)\left(\Box-2H^2
\right)^2h^\sigma_\tau}.  ~~~~~~
\ea      
\section{Adding a cosmological constant}
It should be noted that it is possible to incorporate 
a cosmological constant $\Lambda$ to the linearised equations
of motion by
taking the ``$\Lambda$-gauge'' 
$\partial^\nu h_{\mu\nu}=\frac{1}{2}\partial_\mu h
- \Lambda x_\mu$ \cite{Bernabeu:2011if}. This adds an extra 
term $\Lambda h_{\mu\nu}$ onto the right hand side of 
\eqref{eq:linaeriseomsinddgauge}. This gives us possibilities for
future work.
\section{Other identities}
We require the identities for integrating over a sphere \cite{Carroll:2004st}
\ba \label{eq:integralidentities}
        \int&& d\Omega = 4\pi, \quad 
        \int n_i n_j d\Omega =\frac{4\pi}{3} \delta_{ij}, \non\\
        \int&& n_i n_j n_k n_l d\Omega =\frac{4\pi}{15} \left( \delta_{ij} \delta_{kl} 
        + \delta_{ik}\delta_{jl} + \delta_{il} \delta_{jk}\right),
\ea
\section{Elliptical orbits}
Using our change of coordinates, the integral \eqref{eq:xxcomponentfirstint}
becomes
\ba
       \hat{J}_{xx}= -\frac{\mu}{M} \int^{-\infty}_{\infty} dz ~e^{-z^2 } 
        \frac{\cos^2(\psi( t_r-\frac{z}{M}))
        -\frac{1}{3}}{\left(1+e\cos(\psi (t_r-\frac{z}{M}))\right)^2}.~~~~~~~
\ea 
We can use a Taylor expansion in $\frac{1}{M}$ to write this as the GR expression $J_{xx}$ (the zeroeth order) plus the first order
expression (which disappears as the integrand is odd) 
and finally the second order correction. We use the identities
\ba \label{eq:evenoddidentities}
        \int^{-\infty}_\infty && e^{-z^2}dz=-\sqrt{\pi},
        ~~\quad \int^{-\infty}_\infty e^{-z^2}zdz=0, \non\\
        &&~~~~\int^{-\infty}_\infty e^{-z^2}z^2 dz=-\frac{\sqrt{\pi}}{2},
\ea
to find
\begin{widetext}
\ba \label{eq:expansionofhatJxx}
        \hat{J}_{xx}&\approx& J_{xx}
        -  \frac{\sqrt{\pi}\mu}{24M^2(1+e \cos (\psi))^4}   
          \bigg\{ \psi'^2 \bigg(4 \left(e^2-3\right) \cos (2 \psi)
                -8 e^2-19 e \cos (\psi)+3 e \cos (3 \psi)\bigg)\non\\
        &&-2 \psi'' \sin (\psi) \left(2 \left(e^2+3\right) \cos (\psi)+e (3 \cos (2 \psi)+5)\right)\bigg\}~~~~~~
\ea
We perform a similar calculation for $\hat{J}_{yy}$ to find that 
the full enhancement factor for the IDG term $f^{\text{IDG}}(e)$ 
is given by 
\ba \label{eq:fullidgenhancementfactor}
        f(e)^{\text{IDG}}&=&1-\frac{(9299+111168 \pi ) e}{12282+155520 \pi }
        +\frac{(753298+4783383 \pi ) e^2}{18423+233280 \pi }
        +\frac{(1347719-15413436 \pi ) e^3}{147384+1866240 \pi }
        +\frac{(152362163+521885160 \pi ) e^4}{294768+3732480 \pi }\non\\
        &&-\frac{(6051611+36789444 \pi ) e^5}{72 (2047+25920 \pi )}
        +\frac{(666697961+1567922058 \pi ) e^6}{294768+3732480 \pi }
        -\frac{15 (1908618+1108133 \pi ) e^7}{32752+414720 \pi }\non\\
        && +\frac{(344524449+556982911 \pi ) e^8}{65504+829440 \pi }
        -\frac{(5826870871+2360357712 \pi ) e^9}{1152 (2047+25920 \pi )}
        +\frac{(37373085170+45561968109 \pi ) e^{10}}{4716288+59719680 \pi }\non\\
        &&-\frac{(45892881151+15257013132 \pi ) e^{11}}{6144 (2047+25920 \pi )}
        +\frac{(685593299971+742716547416 \pi ) e^{12}}{36864 (2047+25920 \pi )}
        -\frac{(18923346001+5812048566 \pi ) e^{13}}{2304 (2047+25920 \pi )}\non\\
        && +\frac{(1406663203279+1486964224080 \pi ) e^{14}}{73728 (2047+25920 \pi )}
        -\frac{(612225325649+186007875390 \pi ) e^{15}}{73728 (2047+25920 \pi )}\non\\
        &&+\frac{(1879563787501+1982636168004 \pi ) e^{16}}{98304 (2047+25920 \pi )}
        -\frac{5 (108886731499+33068066736 \pi ) e^{17}}{65536 (2047+25920 \pi )}\non\\
        &&+\frac{(7518767717389+7930544672016 \pi ) e^{18}}{393216 (2047+25920 \pi )}
        -\frac{(9799832804557+2976126006240 \pi ) e^{19}}{1179648 (2047+25920 \pi )}\non\\
        &&
        +\frac{(15037546015045+15861089344032 \pi ) e^{20}}{786432 (2047+25920 \pi )}+ O(e^{21})
\ea
\end{widetext} 
\iffalse
\ba \label{eq:fullidgenhancementfactor}
        f^{\text{IDG}}(e)
        &=&1-\frac{120467e}{167802}+\frac{5284978e^2}{251703}
        -\frac{12620113e^3}{2013624} +\frac{585660427e^4}{4027248}
        -\frac{14387669e^5}{1006812}+\frac{21671843e^6}{55934}
        -\frac{321579275e^7}{4027248}\non\\&&
        +\frac{1822163101e^8}{4027248}
        -\frac{4929137503e^9}{32217984}+\frac{18026523359e^{10}}{64435968}
        -\frac{52454025521e^{11}}{515487744}+\frac{101348994923e^{12}}{1030975488}
        -\frac{14433473687e^{13}}{515487744}\non\\&&+\frac{37007732585e^{14}}{2061950976}
        -\frac{2233524965e^{15}}{687316992}+\frac{12090157079e^{16}}{8247803904}
        -\frac{2100312263e^{17}}{16495607808}+\frac{170855795e^{18}}{3665690624}
        -\frac{26969647e^{19}}{32991215616}\non\\&&+\frac{10580267e^{20}}{21994143744}
        -\frac{11e^{21}}{64435968}+\frac{595e^{22}}{515487744}   
\ea     
\fi 

\newpage 
\bibliographystyle{unsrt}  %use the plain bibliography style
\bibliography{quadrupole}        %use a bibtex bibliography file thesis.bib
\end{document}